\documentclass[aps,twocolumn,superscriptaddress,showpacs,pra,amssymb, amsmath]{revtex4}
\newtheorem{defi}{Definition}
\newtheorem{lem}[defi]{Lemma}
\newtheorem{thm}[defi]{Theorem}

\newtheorem{rem}{Remark}
\def\QED{\mbox{\rule[0pt]{1.5ex}{1.5ex}}}

\def\endproof{\hspace*{\fill}~\QED\par\endtrivlist\unskip}

\def\Tr{\mathop{\rm Tr}\nolimits}

\def\Id{\mathop{I}\nolimits}
\def\rank{\mathop{\rm rank}\nolimits}

\def\real{\mathbb{R}}

\def\SU{\mathop{\rm SU}\nolimits}
\def\Label#1{\label{#1}\ [\ #1\ ]\ }
\def\Bibitem#1{\bibitem{#1}\ [\ #1\ ]\ }
\def\Label{\label}\def\Bibitem{\bibitem}
\begin{document}
\bibliographystyle{unsrt}
\title{Exponents of
quantum fixed-length pure state source coding}
\author{Masahito Hayashi}
\email{masahito@brain.riken.go.jp}
\affiliation{Laboratory for Mathematical Neuroscience, Brain Science Institute, RIKEN,
2-1 Hirosawa, Wako, Saitama, 351-0198, Japan}
\date{12 May 2002}
\begin{abstract}
We derive the optimal exponent of the error probability
of the quantum fixed-length pure state source coding
in both cases of blind coding and visible coding.
The optimal exponent is universally attained by 
Jozsa et al. (PRL, {\bf 81}, 1714 (1998))'s universal code.
In the direct part, a group representation theoretical 
type method is essential.
In the converse part, Nielsen and Kempe
(PRL, {\bf 86}, 5184 (2001))'s lemma
is essential.
\end{abstract}
\maketitle

\section{Introduction}
As was proven by Schumacher \cite{Schumacher}, and
Jozsa and Schumacher \cite{JS},
we can compress the unknown source state
into the coding length $n H(\overline{\rho}_p)$
with a sufficiently small error
when the source state on $n$ quantum systems 
obeys the $n$-independent identical distribution (i.i.d.)
of the known probability $p$,
where $\overline{\rho}_p:= \sum_{\rho} p(\rho) \rho$
and $H(\rho)$ is the von Neumann entropy $- \Tr \rho
\log \rho$.
Jozsa and Schumacher's protocol depends on the mixture state
$\overline{\rho}_p$,
and in this protocol,
the coding length is independent for the input.
Therefore, this type code is called a {\it quantum fixed
length source code}.

Concerning the quantum source coding,
there are two criteria:
One is the blind coding,
in which the input is an unknown quantum state.
The other is the visible coding,
in which the input is classical information
that determines the quantum state,
i.e., the encoder knows the input quantum state.
When a source consists of pure states
and depends on an i.i.d. distribution
of the probability $p$,
the bound of the compression rate
(i.e. the minimum admissible rate)
equals the entropy rate $H(\overline{\rho}_p)$.
The proof of this statement
is divided into two parts:
One is the possibility
to compress the quantum source
into a larger rate than the entropy rate,
which is called the {\it direct part}.
The other is the impossibility
to compress the quantum source
into a smaller rate than the entropy rate,
which is called the {\it converse part}.
The former is given by Schumacher's result.
The latter was proven by Barnum et al.
\cite{Barnum} only in the blind case,
however Horodecki \cite{Horo} proved 
it in both cases by a simpler method.
Winter\cite{Winter} 
proved that the both settings have the 
{\it strong converse} property,
i.e. if we compress into a smaller
rate than the entropy rate,
the average error goes to $1$.
Moreover, depending only on the coding length $n R$,
Jozsa et al. \cite{JH}
constructed a code
which is independent of 
the distribution which the input obeys.
In their protocol,
the average error tends to $0$
when $H(\rho) \,< R$.
Such a code is called 
a {\it quantum universal fixed-length source code}.
Of course, we can consider 
a quantum variable-length source code,
but discuss it in another paper \cite{HayaMa}.

However, only with the knowledge of 
the minimum admissible rate
we cannot estimate 
what a compression rate is available for a given error $\delta \,> 0$
and a given integer $n$.
For such an estimate, we need to discuss
the decreasing speed of the average error for
a fixed rate $R$.
In the classical information theory,
in order to treat this speed, we focus the exponential
rate (exponent) of the error probability,
and the optimal exponent is greater than zero
when the coding rate $R$ is greater than the entropy rate.
Conversely, when the rate $R$ is smaller than
the entropy rate,
the correct probability exponentially goes to zero.
These optimal exponents
have been already calculated by using type method.
(see Csisz\'{a}r and K\"{o}rner \cite{CK}).

In this paper,
we treat only a quantum fixed-length code
at both criteria in the case where
any source consists of pure states.
We optimize the exponents of
the average error and the average fidelity in sec. \ref{s3}.
Using a group representation theoretical type method
introduced in Appendix \ref{appen2}, 
we derive an upper bound of the error
of the quantum universal fixed-length source code
constructed by Jozsa et al. for any $n$ and 
any $R$ as 
(\ref{12-7}), (\ref{12-8}) and (\ref{12-9})
in sec. \ref{s4}.
This upper bound yields its attainability
of the optimal exponents.
In sec. \ref{s6}, non-existence of 
a code exceeding the exponents
is proven, which is called
the converse part.
In the converse part, an inequality is essential
and is proven from Nielsen and Kempe's lemma \cite{NK}
in sec. \ref{s5}.
\section{Summary of previous results}
Blind and visible codes are mathematically formulated
as follows.
Assume that
a quantum pure state $\rho_i$ on ${\cal H}$ corresponding to
label $i \in \Xi$ 
is generated with probability $p_i$ .
We denote the set of quantum states
on ${\cal H}$ by ${\cal S}({\cal H})$.
Therefore, the source is described by 
$\{\rho_i,p_i\}_{i \in  \Xi}$.
In the blind setting,
the encoder is described by 
a CP map $E$ from ${\cal S}({\cal H})$
to ${\cal S}({\cal K})$,
and the decoder is described by a CP map $D$ from ${\cal S}({\cal K})$
to ${\cal S}({\cal H})$.
The average error 
is given by 
$\epsilon(E,D):= \sum_{i \in \Xi} p_i 
(1 - \Tr D \circ E (\rho_i) \rho_i)$,
and the average fidelity 
is given by $\sum_{i \in \Xi} p_i \Tr D \circ E (\rho_i) \rho_i$.
We call a triple $({\cal K},E,D)$
a blind code.

In the visible setting,
the encoder is described by 
a map $F$ from $\Xi$
to ${\cal S}({\cal K})$.
Then, the average error 
is given by 
$\epsilon(F,D):= \sum_{i \in \Xi} p_i 
(1 - \Tr D \circ F(i) \rho_i)$.
In this setting, we treat the trade-off
between decreasing 
$\dim {\cal K}$ and $\epsilon(F,D)$.
We call a triple $({\cal K},E,D)$
a blind code.
Similarly, we call a triple $({\cal K},F,D)$
a visible code.
In the both settings, we treat the trade-off
between decreasing 
$\dim {\cal K}$ and 
$\epsilon(E,D)$
($\epsilon(F,D)$).

A blind code $({\cal K},E,D)$ can be regarded
as a visible code
in the case
where $F(i):= E( \rho_i)$.
We have more choices in the visible setting
than in the blind setting.
A blind code is used for saving memories in 
quantum computing.
A visible code is used for efficient use of 
quantum channel in quantum cryptography,
for example, the B92 protocol
\cite{B92}, \cite{BCFJS}.

In the $n$-i.i.d. setting,
the quantum state $\rho_{n, \vec{i}_n}:=
\rho_{i_1} \otimes \rho_{i_2} \otimes \cdots\otimes\rho_{i_n}$
on the tensored Hilbert space
${\cal H}^{\otimes n}$
generates with the probability
$p_{n, \vec{i}_n}:=
p_{i_1}p_{i_2} \cdots p_{i_n}$,
where $\vec{i}_n= (i_1, i_2, \ldots, i_n)$.
This setting is written by
the source 
$\{ \rho_{n, \vec{i}_n},p_{n, \vec{i}_n}\}_{\vec{i}_n\in \Xi^n}$,
which is called 
a $n$-{\it discrete memoryless source} (DMS)
generated by the source $\{\rho_i,p_i\}_{i \in  \Xi}$.
Now, we define
the minimum admissible rate 
$R_B(\{\rho_i,p_i\}_{i \in  \Xi})$
($R_V(\{\rho_i,p_i\}_{i \in  \Xi})$)
and the converse minimum admissible rate 
$R_B^-(\{\rho_i,p_i\}_{i \in  \Xi})$
($R_V^-(\{\rho_i,p_i\}_{i \in  \Xi})$)
of the DMS generated by 
$\{\rho_i,p_i\}_{i \in  \Xi}$
in the blind setting (in the visible setting)
as follows,
respectively.
\begin{align*}
&R_B(\{\rho_i,p_i\}_{i \in  \Xi}) \\
:= 
&\inf\left\{\left.
\varlimsup
\frac{1}{n}\log \dim {\cal K}_n\right|
\begin{array}{ll}
\exists \{ ({\cal K}_n,E_n,D_n)\},\\
\epsilon(E_n,D_n) \to 0 
\end{array}
\right \} \\
&R_V(\{\rho_i,p_i\}_{i \in  \Xi})\\
:=
&\inf\left\{\left.
\varlimsup
\frac{1}{n}\log \dim {\cal K}_n\right|
\begin{array}{ll}
\exists \{ ({\cal K}_n,F_n,D_n)\},\\
\epsilon(F_n,D_n) \to 0 
\end{array}
\right \}\\
&R_B^-(\{\rho_i,p_i\}_{i \in  \Xi})\\
:=
&\inf\left\{\left.
\varlimsup
\frac{1}{n}\log \dim {\cal K}_n\right|
\begin{array}{ll}
\exists \{ ({\cal K}_n,E_n,D_n)\},\\
\varlimsup  \epsilon(E_n,D_n) \,< 1 
\end{array}
\right \} \\
&R_V^-(\{\rho_i,p_i\}_{i \in  \Xi})\\
:=
&\inf\left\{\left.
\varlimsup
\frac{1}{n}\log \dim {\cal K}_n\right|
\begin{array}{ll}
\exists \{ ({\cal K}_n,F_n,D_n)\},\\
\varlimsup
\epsilon(F_n,D_n) \,< 1 
\end{array}
\right \}.
\end{align*}

The following theorem is a known result.
\begin{thm}\Label{T1}
The equations
\begin{align}
&R_B(\{\rho_i,p_i\}_{i \in  \Xi})
=
R_V(\{\rho_i,p_i\}_{i \in  \Xi})
=R_B^-(\{\rho_i,p_i\}_{i \in  \Xi}) \nonumber \\
&=
R_V^-(\{\rho_i,p_i\}_{i \in  \Xi})
= H(\overline{\rho}_p) \Label{1}
\end{align}
hold, where
$\overline{\rho}_p:= \sum_{i \in \Xi} p_i \rho_i$ and
$H(\rho)$ denotes von Neumann entropy 
$-\Tr \rho\log \rho$.
\end{thm}
Since the following relations
\begin{align*}
& R_B(\{\rho_i,p_i\}_{i \in  \Xi})
\ge R_V(\{\rho_i,p_i\}_{i \in  \Xi}),\\
& R_B^-(\{\rho_i,p_i\}_{i \in  \Xi})
\ge R_V^-(\{\rho_i,p_i\}_{i \in  \Xi}),\\
& R_B(\{\rho_i,p_i\}_{i \in  \Xi}) \ge
R_B^-(\{\rho_i,p_i\}_{i \in  \Xi}), \\
& R_V(\{\rho_i,p_i\}_{i \in  \Xi}) \ge
R_V^-(\{\rho_i,p_i\}_{i \in  \Xi})
\end{align*}
are trivial,
it is sufficient for (\ref{1}) to prove 
\begin{align*}
R_B(\{\rho_i,p_i\}_{i \in  \Xi})
\le H(\rho) , \quad
R_V^-(\{\rho_i,p_i\}_{i \in  \Xi})
\ge H(\rho).
\end{align*}

Schumacher \cite{Schumacher} proved the direct part:
$R_B(\{\rho_i,p_i\}_{i \in  \Xi})
\le H(\overline{\rho}_p)$, and
Jozsa-Schumacher \cite{JS} simplified it.
Barnum et al. \cite{Barnum} proved 
the weak converse part: $R_B(\{\rho_i,p_i\}_{i \in  \Xi})
\ge H(\overline{\rho}_p)$ of the blind case,
and 
Horodecki \cite{Horo} proved
the weak converse part: $R_V(\{\rho_i,p_i\}_{i \in  \Xi})
\ge H(\overline{\rho}_p)$ of the visible case,
which is a stronger argument than
the one of the blind case.
Winter \cite{Winter} obtained the 
strong converse part:
$R_V^-(\{\rho_i,p_i\}_{i \in  \Xi})
\ge H(\overline{\rho}_p)$.
Moreover, Petz and Mosonyi \cite{PM}
treated the general stationary case,
in which there are memory effects.

\section{Main results}\Label{s3}
Next, we define the exponents of the average error
(the reliable functions)
$r_{e,B}(R|\{\rho_i,p_i\}_{i \in  \Xi})$ and
$r_{e,V}(R|\{\rho_i,p_i\}_{i \in  \Xi})$,
and the exponents of the average fidelity
(the converse reliable functions)
$r_{e,B}^*(R|\{\rho_i,p_i\}_{i \in  \Xi})$ and 
$r_{e,V}^*(R|\{\rho_i,p_i\}_{i \in  \Xi})$ by
\begin{align*}
& r_{e,B}(R|\{\rho_i,p_i\}_{i \in  \Xi}) \\
:=
& \sup \left\{ \varliminf \frac{-1}{n}\log\epsilon(E_n,D_n) 
\left|
\begin{array}{l}
\exists \{ ({\cal K}_n,E_n,D_n)\},\\
\varlimsup \frac{1}{n}\log \dim {\cal K}_n  \le R 
\end{array}
\right.\right\} \\
&r_{e,V}(R|\{\rho_i,p_i\}_{i \in  \Xi})\\
:= 
&\sup \left\{ \varliminf \frac{-1}{n}\log\epsilon(F_n,D_n) 
\left|
\begin{array}{l}
\exists \{ ({\cal K}_n,F_n,D_n)\},\\
\varlimsup \frac{1}{n}\log \dim {\cal K}_n  \le R 
\end{array}
\right.
\right\} \\
&r_{e,B}^*(R|\{\rho_i,p_i\}_{i \in  \Xi})\\
:=
&\inf \left\{ \varlimsup \frac{-1}{n}\log (1- \epsilon(E_n,D_n) )
\left|
\begin{array}{l}
\exists \{ ({\cal K}_n,E_n,D_n)\},\\
\varlimsup \frac{1}{n}\log \dim {\cal K}_n  \le R 
\end{array}
\right.\right\} \\
& r_{e,V}^*(R|\{\rho_i,p_i\}_{i \in  \Xi})\\
:= 
& \inf \left\{\varlimsup \frac{-1}{n}\log(1-\epsilon(F_n,D_n) )
\left|
\begin{array}{l}
\exists \{ ({\cal K}_n,F_n,D_n)\},\\
\varlimsup \frac{1}{n}\log \dim {\cal K}_n  \le R 
\end{array}
\right.\right\} .
\end{align*}

The following is the main theorem.
\begin{thm}\Label{T20}
Assume that $0 \le R \,< \log d$ and 
$d= \dim {\cal H}$.
We diagonalize $\overline{\rho}_p$ as
$\overline{\rho}_p=\sum_i a_i |e_i\rangle
\langle e_i|$ such that $a_i\ge a_{i+1}$.
Then, ${\bf a}:= \{a_i\}$ is
a probability distribution on $\{ 1, \ldots, d\}$.
The relations 
\begin{align}
r_{e,B}(R|\{\rho_i,p_i\}_{i \in  \Xi})
&= r_{e,V}(R|\{\rho_i,p_i\}_{i \in  \Xi})\nonumber \\
&= \max_{0 \,< s \le 1} \frac{(1-s)R - \psi(s)}{s} \Label{T2} \\
&= \min_{H( \sigma) \ge R } D(\sigma\| \overline{\rho}_p) 
 \Label{T2.1} \\
&= \min_{ H( {\bf b}) \ge R } D( {\bf b}\| {\bf a})
 \Label{T2.11} \\
r_{e,B}^*(R|\{\rho_i,p_i\}_{i \in  \Xi}) &\ge
r_{e,V}^*(R|\{\rho_i,p_i\}_{i \in  \Xi}) 
\nonumber\\
& = \sup_{s \ge 1} \frac{(1-s)R - \psi(s)}{s}  \Label{T4} \\
& = \min_{H( \sigma) \le R } D(\sigma\| \overline{\rho}_p) 
\Label{T4.1}\\ 
& = \min_{H( {\bf b}) \le R } D({\bf b}\|{\bf a}) 
\Label{T4.11} 
\end{align}
hold, where
$\psi(s)$ denotes the R\'{e}ny entropy $\log \Tr \overline{\rho}_p^{s}$,
$D(\sigma \| \rho)$ denotes the quantum 
relative entropy $\Tr \sigma (\log \sigma - \log\rho)$,
and ${\bf b}$ denotes a probability on $\{ 1, \ldots, d\}$.
\end{thm}
Our proof of Theorem \ref{T20}
is outlined as follows.
Since any blind code can be demonstrated as
a visible code,
the relations
\begin{align}
r_{e,B}(R|\{\rho_i,p_i\}_{i \in  \Xi})
&\le r_{e,V}(R|\{\rho_i,p_i\}_{i \in  \Xi}) \\
r_{e,B}^*(R|\{\rho_i,p_i\}_{i \in  \Xi}) 
&\ge
r_{e,V}^*(R|\{\rho_i,p_i\}_{i \in  \Xi}) 
\end{align}
are trivial.
In sec.\ref{s4},
we universally construct the optimal quantum 
fixed-length code with the rate $R$.
This construction is independent of 
$\overline{\rho}_p$,
and depends only on the rate $R$.
From this construction, we obtain 
\begin{align}
r_{e,B}(R|\{\rho_i,p_i\}_{i \in  \Xi})
&\ge  \min_{ H( {\bf b}) \ge R } 
D( {\bf b}\|{\bf a}) \Label{12-1} \\
r_{e,V}^*(R|\{\rho_i,p_i\}_{i \in  \Xi}) 
& \le \min_{H( {\bf b}) \le R } D({\bf b}\|{\bf a}) ,\Label{12-2}
\end{align}
which is called the direct part.
In sec. \ref{s6},
we prove
\begin{align}
r_{e,V}(R|\{\rho_i,p_i\}_{i \in  \Xi})
&\le  \max_{0 \,< s \le 1} 
\frac{(1-s)R - \psi(s)}{s}\Label{12-3}\\
r_{e,V}^*(R|\{\rho_i,p_i\}_{i \in  \Xi}) 
& \ge \sup_{s \ge 1} \frac{(1-s)R - \psi(s)}{s}
\Label{12-4},
\end{align}
which is called the converse part.
The equivalence between RHSs of 
(\ref{12-1}),(\ref{12-3}) and (\ref{T2.1})
((\ref{12-2}),(\ref{12-4}) and (\ref{T4.1}))
is proven in Appendix \ref{appen1},
respectively.

\begin{rem}
The inequality $r_{e,V}(R|\{\rho_i,p_i\}_{i \in  \Xi})\ge
\min\{ D(\sigma\| \rho) | H( \sigma) \le R \}$ 
was proven by Winter \cite{Winter}.
\end{rem}
\begin{rem}
We can adopt another criteria
for error as:
\begin{align*}
\epsilon_b(E,D)&:= \sum_{i \in \Xi} p_i 
(1 - \sqrt{\Tr D \circ F (\rho_i) \rho_i}) \\
\epsilon_b(F,D)&:= \sum_{i \in \Xi} p_i 
(1 - \sqrt{\Tr D \circ F(i) \rho_i}).
\end{align*}
Note that
$(1 - \sqrt{\Tr D \circ F (i) \rho_i})
=
(1 - \Tr | D \circ F (i) \rho_i|)$
equals Bures distance.
In this case, we can define other reliable functions
$r_{e,B,b}(R|\{\rho_i,p_i\}_{i \in  \Xi})$ and
$r_{e,V,b}(R|\{\rho_i,p_i\}_{i \in  \Xi})$,
and other converse reliable functions
$r_{e,B,b}^*(R|\{\rho_i,p_i\}_{i \in  \Xi})$ and 
$r_{e,V,b}^*(R|\{\rho_i,p_i\}_{i \in  \Xi})$ by
\begin{align*}
& r_{e,B,b}(R|\{\rho_i,p_i\}_{i \in  \Xi})\\
:=
& \sup \left\{ \varliminf \frac{-1}{n}\log \epsilon_b(E_n,D_n) 
\left|
\begin{array}{l}
\exists \{ ({\cal K}_n,E_n,D_n)\},\\
\varlimsup \frac{1}{n}\log \dim {\cal K}_n  \le R 
\end{array}
\right.\right\} \\
& r_{e,V,b}(R|\{\rho_i,p_i\}_{i \in  \Xi})\\
:= 
&\sup \left\{ \varliminf \frac{-1}{n}\log \epsilon_b(F_n,D_n) 
\left|
\begin{array}{l}
\exists \{ ({\cal K}_n,F_n,D_n)\},\\
\varlimsup \frac{1}{n}\log \dim {\cal K}_n  \le R 
\end{array}
\right.
\right\} \\
&r_{e,B,b}^*(R|\{\rho_i,p_i\}_{i \in  \Xi})\\
:=
&\inf \left\{ \varlimsup \frac{-1}{n}\log (1- \epsilon_b(E_n,D_n) )
\left|
\begin{array}{l}
\exists \{ ({\cal K}_n,E_n,D_n)\},\\
\varlimsup \frac{1}{n}\log \dim {\cal K}_n  \le R 
\end{array}
\right.\right\} \\
& r_{e,V,b}^*(R|\{\rho_i,p_i\}_{i \in  \Xi})\\
:= 
& \inf \left\{\varlimsup \frac{-1}{n}\log(1-\epsilon_b(F_n,D_n) )
\left|
\begin{array}{l}
\exists \{ ({\cal K}_n,F_n,D_n)\},\\
\varlimsup \frac{1}{n}\log \dim {\cal K}_n  \le R 
\end{array}
\right.\right\} .
\end{align*}
As proven in Appendix \ref{apec},
the following relations between two criteria
\begin{align}
r_{e,B,b}(R|\{\rho_i,p_i\}_{i \in  \Xi})
&=r_{e,B}(R|\{\rho_i,p_i\}_{i \in  \Xi}) \Label{h17}\\
r_{e,V,b}(R|\{\rho_i,p_i\}_{i \in  \Xi})
&=r_{e,V}(R|\{\rho_i,p_i\}_{i \in  \Xi}) \Label{h16}\\
r_{e,B}^*(R|\{\rho_i,p_i\}_{i \in  \Xi})
& \ge r_{e,B,b}^*(R|\{\rho_i,p_i\}_{i \in  \Xi}) \Label{h27}\\
r_{e,V}^*(R|\{\rho_i,p_i\}_{i \in  \Xi})
& \ge r_{e,V,b}^*(R|\{\rho_i,p_i\}_{i \in  \Xi}) \Label{h26}
\end{align}
hold.
\end{rem}
\section{Construction of a universal fixed-length source 
code to achieve the optimal rate}\Label{s4}
We construct
a universal quantum fixed-length source code
to achieve the optimal rate
in Theorem \ref{T20}.
For any $r \,> 0$ and $R\,> 0$,
the set $\{ \rho \in {\cal S}({\cal H})|
\min_{H(\sigma) \ge R} D(\sigma\|\rho) = r \}$
is covariant
for the actions of the $d$-dimensional special unitary group $\SU(d)$,
and any $n$-i.i.d. distribution $p^n$
is invariant for the action of the $n$-th 
symmetric group $S_n$
on the tensored space ${\cal H}^{\otimes n}$.
Thus, our code should satisfy the invariance
for these actions on ${\cal H}^{\otimes n}$.

Now, we focus on the irreducible decomposition 
of the tensored space ${\cal H}^{\otimes n}$
concerning the representations of 
$S_n$ and $\SU(d)$,
and define the Young index ${\bf n}$ as,
\begin{align*}
{\bf n}: = (n_1, \ldots, n_d) , \quad \sum_{i=1}^d
n_i= n, n_{i} \ge n_{i+1},
\end{align*}
and denote the set of 
Young indices ${\bf n}$ by $Y_n$.
Young index ${\bf n}$ uniquely corresponds to 
the irreducible unitary representation 
of $S_n$ and
the one of $\SU(d)$.
Now, we denote the representation space
of the irreducible unitary representation 
of $S_n$ ($\SU(d)$) corresponding to ${\bf n}$
by ${\cal V}_{{\bf n}}$ (${\cal U}_{{\bf n}}$), respectively.
In particular, regarding a unitary representation 
of $\SU(d)$,
Young index ${\bf n}$ gives the highest weight of 
the corresponding representation.
Then, the tensored space ${\cal H}^{\otimes n}$ is 
decomposed as follows; i.e.
${\cal H}^{\otimes n}$ is equivalent with the following
direct sum space under the representation of
$S_n$ and $\SU(d)$.
\begin{align*}
{\cal H}^{\otimes n}=
\bigoplus_{{\bf n}}{\cal W}_{{\bf n}} , \quad
{\cal W}_{{\bf n}}:= {\cal U}_{{\bf n}} \otimes {\cal V}_{{\bf n}} .
\end{align*}
Since this representation of the group $S_n \times \SU(d)$
is unitary,
any irreducible components ${\cal W}_{\bf n}$ are
orthogonal with one another.
For details, see Weyl \cite{Weyl}, 
Goodman and Wallach \cite{GW}, and Iwahori \cite{Iwa}.
The efficiency of this representation method
was discussed from several viewpoints.
Regarding fixed-length source coding,
it was discussed by
Jozsa et. al. \cite{JH}.
Regarding quantum relative entropy,
it was by Hayashi\cite{H1997}.
Regarding quantum hypothesis testing,
it was by Hayashi\cite{H2001}.
Regarding estimation of spectrum,
it was by Keyl and Werner\cite{KW}.

Next, we construct a blind code with rate $R$.
We define the Hilbert space ${\cal K}_{R,n}$,
the blind encoder $E_{R,n}$,
the visible encoder $F_{R,n}$ and 
the decoder $D_{R,n}$ 
by
\begin{align*}
{\cal K}_{R,n}:=&\bigoplus_{{\bf n}:H(\frac{\bf n}{n}) \le R}
{\cal W}_{{\bf n}} \\
E_{R,n}(\rho):=&
P_{R,n} \rho P_{R,n} + \Tr \rho (I-P_{R,n})
\frac{I_{{\cal K}_{R,n}}}{\Tr I_{{\cal K}_{R,n}}} \\
F_{R,n}(\vec{i}_n):=&
\frac{P_{R,n} \rho_{n,\vec{i}_n} P_{R,n} }
{\Tr P_{R,n} \rho_{n,\vec{i}_n} P_{R,n} } \\
D_{R,n}(\rho):=&
\rho ,
\end{align*}
where 
we denote the projection to ${\cal K}_{R,n}$ by
$P_{R,n}$.

\begin{lem}\Label{L3}
We define $R_n$ by
\begin{align}
R_n:= R- \frac{4d}{n}\log (n+d).
\end{align}
The rates of the blind code $\{ ({\cal K}_{R,n},E_{R,n},D_{R,n})\}$
and the visible code $\{ ({\cal K}_{R,n},F_{R,n},D_{R,n})\}$
satisfies 
\begin{align}
\dim {\cal K}_{R_n,n}
\le e^{n R}.  \Label{10-1}
\end{align}
When the mixture $\overline{\rho}_p$ of the source 
is diagonalized as $\sum_{j=1}^d
a_j|e_j\rangle \langle e_j|$,
we can evaluate the average errors as 
\begin{align}
&\epsilon (F_{R_n,n},D_{R_n,n})\nonumber \\
\le& (n+d)^{4d}
\exp\left(- n \min_{H({\bf b})\ge R_n}
D({\bf b}\|{\bf a}) \right)\Label{12-7}\\
&\epsilon (E_{R_n,n},D_{R_n,n})\nonumber \\
\le& 2
 (n+d)^{4d}
\exp\left(- n \min_{H({\bf b})\ge R_n} 
D({\bf b}\|{\bf a})\right)\Label{12-8}\\
&1- \epsilon (F_{R_n,n},D_{R_n,n})\nonumber \\
\ge&
(n+d)^{-\frac{d(d+1)}{2}}
\exp\left( -n 
\min_{{\bf n}\in Y_n: H\left(\frac{\bf n}{n}\right)
\le R_n}D\left(\left.\frac{\bf n}{n}
\right\|{\bf a}\right)
\right),\Label{12-9}
\end{align}
where 
${\bf a}$ is defined as ${\bf a}:=\{a_i\}$ and 
${\bf b}=\{b_i\}$ denotes a probability on 
$\{ 1, \ldots, d\}$.
Taking the limit, we obtain
\begin{align}
\varliminf \frac{-1}{n}\log \epsilon (E_{R_n,n},D_{R_n,n})
&\ge 
\min_{H({\bf b})\ge R}D({\bf b}\|{\bf a}) , \Label{12-5}\\
\varlimsup \frac{-1}{n}\log \left(
1- \epsilon (F_{R_n,n},D_{R_n,n})\right)
&\le 
\min_{H({\bf b})\le R}D({\bf b}\|{\bf a}) .\Label{12-6}
\end{align}
\end{lem} 
Inequalities (\ref{12-5}) and (\ref{12-6})
imply (\ref{12-1}) and (\ref{12-2}),
respectively.
Conversely, the opposite inequalities of
(\ref{12-5}) and (\ref{12-6})
are guaranteed by inequalities (\ref{12-3}) and
(\ref{12-4}).
\begin{rem}
The subspace ${\cal K}_{R_n,n}$ is equal to the subspace 
$\Upsilon$ introduced by Jozsa et al. \cite{JH}
because both are invariant for the action of the
symmetric group.
Therefore, our code $E_{R_n,n}$ coincides with
their protocol.
\end{rem}

\begin{rem}
Even if the source states $\rho_i$ are not
pure, 
we can prove inequalities similar 
to (\ref{12-7}), (\ref{12-8}) and (\ref{12-9})
by using some calculations similar to
Appendix C in Hayashi and Matsumoto\cite{HayaMa}.
However, in this case,
this exponent does not seem to be optimal.
\end{rem}

\noindent\hspace{2em}{\it Proof of Lemma \ref{L3}: }
Using Lemma \ref{L110},
we can evaluate as
\begin{align*}
\dim {\cal K}
&\le (n +1)^d 
\max_{{\bf n}\in Y_n: H\left(\frac{\bf n}{n}\right)
\le R_n} \dim {\cal W}_{\bf n} \\
&\le (n+1)^{2d}
\max_{{\bf n}\in Y_n: H
\left(\frac{\bf n}{n}\right)\le R_n} 
\dim {\cal V}_{\bf n} \\
&\le (n+1)^{2d}(n+d)^{2d}e^{n R_n}.
\end{align*}
Thus, we obtain (\ref{10-1}).
The average error of the visible code
can be calculated as
\begin{align*}
&\epsilon(F_{R_n,n}, D_{R_n,n})\\
=&
\sum_{\vec{i}_n \in \Xi^n}p_{n,\vec{i}_n}
\left(
1- \Tr \rho_{n,\vec{i}_n} \frac{P_{R_n,n} \rho_{n,\vec{i}_n} P_{R_n,n} }
{\Tr P_{R_n,n} \rho_{n,\vec{i}_n} P_{R_n,n} } \right)\\
=&\sum_{\vec{i}_n \in \Xi^n}p_{n,\vec{i}_n}
\left(1- \Tr P_{R_n,n} \rho_{n,\vec{i}_n}\right)\\
=&
\left(1- \Tr P_{R_n,n} 
\sum_{\vec{i}_n \in \Xi^n}p_{n,\vec{i}_n}\rho_{n,\vec{i}_n}\right)\\
=&
\left(1- \Tr P_{R_n,n} \overline{\rho}_p^{\otimes n}
\right).
\end{align*}
Therefore, Lemma \ref{lee} guarantees (\ref{12-7})
and (\ref{12-9}).
Conversely,
\begin{align*}
&\epsilon(E_{R_n,n}, D_{R_n,n})\\
=&
\sum_{\vec{i}_n \in \Xi^n}p_{n,\vec{i}_n}
\Biggl[
1- \Tr \rho_{n,\vec{i}_n} \\
&
\left( P_{R_n,n} \rho_{n,\vec{i}_n} P_{R_n,n} 
+
\Tr \rho_{n,\vec{i}_n} (I-P_{R_n,n})
\frac{I_{{\cal K}_{R_n,n}}}{\Tr I_{{\cal K}_{R_n,n}}}
\right)
\Biggr] \\
 \le&
\sum_{\vec{i}_n \in \Xi^n}p_{n,\vec{i}_n}
\left(
1- \Tr \rho_{n,\vec{i}_n} P_{R_n,n} \rho_{n,\vec{i}_n} P_{R_n,n} 
\right)\\
=&
\sum_{\vec{i}_n \in \Xi^n}p_{n,\vec{i}_n}
\left(
1- (\Tr \rho_{n,\vec{i}_n} P_{R_n,n} )^2
\right)\\
 \le&
\left(
1- \left(
\sum_{\vec{i}_n \in \Xi^n}p_{n,\vec{i}_n}
\Tr \rho_{n,\vec{i}_n} P_{R_n,n} 
\right)^2
\right)\\
 =&
1- \left(
\Tr \overline{\rho}_{p}^{\otimes n} P_{R_n,n} 
\right)^2 \le
2 \left(1- 
\Tr \overline{\rho}_{p}^{\otimes n} P_{R_n,n} 
\right)
\end{align*}
which implies (\ref{12-8}).
\endproof

\section{Necessary inequality for the converse part}\Label{s5}
For an Hermitian matrix $X$,
we define
the projections $\{ X \ge 0 \},\{ X \,< 0 \}$
by
\begin{align*}
\{ X \ge 0 \} = \sum_{s_j \ge 0 }E_j, \quad
\{ X \,< 0 \}= \sum_{s_j \,< 0 }E_j,
\end{align*}
where the spectral decomposition
of $X$ is given by
$X = \sum_{j} s_j E_j$
($s_j$ is an eigenvalue corresponding to projection $E_j$).
Under a source $\{\rho_i,p_i\}_{i \in  \Xi}$,
the following lemma holds.
\begin{lem}\Label{L1}
Any visible code $({\cal K},F,D)$ 
satisfies the following inequalities
\begin{align}
\epsilon(F,D)+ e^\lambda \dim {\cal K} &\ge
\Tr \overline{\rho}_p \{ \overline{\rho}_p - e^\lambda \,< 0 \} \Label{L10} \\
1- \epsilon(F,D) &\le  e^{\lambda} \dim {\cal K} +
 \Tr \overline{\rho}_p \{ \overline{\rho}_p - e^\lambda \ge 0 \}\Label{L20}
\end{align}
for $\forall \lambda \in \real$.

Moreover, the inequality
\begin{align}
1- \epsilon(F,D) &\le  e^{\lambda} \dim {\cal K} +
e^{(1-s)\lambda + \psi(s)}
 \Label{L21}
\end{align}
holds for $\forall \lambda \in \real, \forall s \ge 1$.
\end{lem}

For our proof of the above lemma,
we require the following two lemmas.
\begin{lem}\Label{L81}
The set of visible encoders
from $\Xi$ to ${\cal S}({\cal K})$
coincides with the convex hull 
of the set of extremal points,
which equals
\begin{align}
\left\{ F \left|
F(i) \hbox{ is a pure state } \forall i \in \Xi\right.
\right\}. \Label{a1}
\end{align}
\end{lem}
\begin{proof}
If a visible encoder $F$
satisfies that
$f(i)$ is a pure state for any $i \in \Xi$,
then $F$ is an extremal point.
It is sufficient to show that
for any visible encoder 
$
F(i)=\sum_{j_i} s_{j_i} |\phi_{j_i}\rangle \langle \phi_{j_i}|
$
is written by a convex hull of (\ref{a1}).
A visible encoder $F(j_1, j_2, \ldots, j_n)$ 
defined by
\begin{align*}
F(j_1, j_2, \ldots, j_n| i)= |\phi_{j_i}\rangle \langle
\phi_{j_i}|
\end{align*}
belongs to (\ref{a1}).
Since the relation 
$F= \sum_{j_1, j_2, \ldots, j_n}
s_{j_1} s_{j_2} \cdots s_{j_n}
F(j_1, j_2, \cdots, j_n)$ holds,
we obtain the lemma.
\end{proof}
\begin{lem}\Label{L82}
The set of decoders
from ${\cal S}({\cal K})$
to ${\cal S}({\cal H})$
coincides with the convex hull 
of the subset 
\begin{align}
\left\{ D \left|
\begin{array}{l}
\hbox{There exists a Hilbert space } 
{\cal H}' 
\hbox{ and} \\
\hbox{an isometry } T 
\hbox{ from }
{\cal S}({\cal K})
\hbox{ to }
{\cal S}({\cal H} \otimes {\cal H}') \\
\hbox{ such that }
D(\rho) = \Tr_{{\cal H}'} T(\rho) .
\end{array}
\right. \right\}. \Label{a2}
\end{align}
\end{lem}
\begin{proof} From the Steinspring representation theorem,
there exist a Hilbert space ${\cal K}'$ 
and a unitary $U$ on 
${\cal K}\otimes {\cal K}'\otimes {\cal H}$
and 
an element $\rho_0 \in {\cal S}({\cal K}'\otimes {\cal H})$ 
such that
\begin{align*}
D(\rho)= \Tr_{{\cal K}\otimes {\cal K}'} U \rho\otimes \rho_0
U^*,\quad \forall \rho \in {\cal S}({\cal K}).
\end{align*}
Assume that
$\rho_0 = \sum_{j} s_j | \phi_j \rangle \langle \phi_j|$.
Then, the decoder $D_j$:
\begin{align*}
D_j(\rho)= \Tr_{{\cal K}\otimes {\cal K}'} U \rho\otimes 
| \phi_j \rangle \langle \phi_j|
U^*,\quad \forall \rho \in {\cal S}({\cal K})
\end{align*}
belongs to (\ref{a2}).
Since $D= \sum_j s_j D_j$,
the proof is complete.
\end{proof}

For a proof of Lemma \ref{L1},
an entanglement viewpoint 
plays a essential role.
A state $\rho \in {\cal S}({\cal H}_A \otimes {\cal H}_B)$
is called separable if
there exist states $\rho_{A,i} \in {\cal S}({\cal H}_A),
\rho_{B,i} \in {\cal S}({\cal H}_B)$
and a probability $p_i$ such that
\begin{align*}
\rho= \sum_i p_i \rho_{A,i} \otimes \rho_{B,i}.
\end{align*}
The following lemma
was proven from the viewpoint of
entanglement by Nielsen and Kempe \cite{NK}.
\begin{lem}\Label{L7}
When the state $\rho \in {\cal S}({\cal H}_A \otimes {\cal H}_B)$
is separable,
the inequality 
\begin{align*}
&\max\{ \Tr P\rho_A | P: \hbox{ projection on } {\cal H}_A, \rank P = k \}
\\
\ge &
\max\{ \Tr P\rho | P: \hbox{ projection on }
{\cal H}_A \otimes {\cal H}_B , \rank P = k \}
\end{align*}
holds for any integer $k$,
where
$\rho_A:= \Tr_{{\cal H}_B} \rho$.
\end{lem}
\noindent\hspace{2em}{\it Proof of Lemma \ref{L1}: }From
Lemma \ref{L81} and Lemma \ref{L82},
it is sufficient to 
show the inequalities (\ref{L10}), (\ref{L20}) and (\ref{L21})
for the pair an encoder $F$ belonging to (\ref{a1})
and a decoder $D$ belonging to (\ref{a2}).
Assume that the Hilbert space ${\cal H}'$
satisfies that $D(\rho)= \Tr_{{\cal H}'}T(\rho)$.
The state $\rho_i':=
\frac{\rho_i \otimes \Id T(F(i))\rho_i \otimes \Id }
{\Tr T(F(i))\rho_i \otimes \Id }\in
{\cal S}({\cal H} \otimes {\cal H}')$
is pure and
satisfies that
$\Tr D(F(i))\rho_i = \Tr T(F(i))\rho_i \otimes \Id
= \Tr  T(F(i))\rho_i '$.
Since $\Tr_{{\cal H}'}\rho_i'= \rho_i$,
there exists a pure state $\sigma_i \in {\cal S}({\cal H}')$
such that
$\rho_i'=\rho_i\otimes \sigma_i$.
Since the state 
$\overline{\rho}_p' := \sum_{i \in \Xi} p_i \rho_i'
= \sum_{i \in \Xi} p_i\rho_i\otimes \sigma_i$
is separable
and $\overline{\rho}_p=\Tr_{{\cal H}'}\overline{\rho}_p'$,
Lemma \ref{L7} guarantees that
\begin{align}
&\max\{ \Tr P \overline{\rho}_p' | P:\hbox{ projection on }
{\cal H}\otimes {\cal H}', \rank P= \dim {\cal K} \} \nonumber\\
\le& \max\{ \Tr P \overline{\rho}_p | P:\hbox{ projection on }
{\cal H}, \rank P= \dim {\cal K} \} \Label{h11}.
\end{align}
Since $\Id \ge F(i)$, we have $T(\Id) \ge  T(F(i))$.
The relations
\begin{align}
& \sum_{i \in \Xi} p_i \Tr D(F(i))\rho_i
= \sum_{i \in \Xi} p_i \Tr  T(F(i))\rho_i' \nonumber\\
& \le \sum_{i \in \Xi} p_i \Tr  T(\Id)\rho_i'
= \Tr  T(\Id)\overline{\rho}_p'
\Label{L24}
\end{align}
hold.
The relations $\Id \ge T(\Id) \ge 0$ and 
$\Tr T(\Id)= \Tr \Id_{{\cal K}}= \dim {\cal K} $
imply that
\begin{align}
\Tr  T(\Id)\overline{\rho}_p'  
\le 
\max\left\{ 
\Tr P \overline{\rho}_p' \left| 
\begin{array}{l}
P:\hbox{ projection on }
{\cal H}\otimes {\cal H}', \\
\rank P= \dim {\cal K} 
\end{array}
\right.\right\}.
\Label{kore}
\end{align}

Assume that $P$ is a projection on ${\cal H}$
whose rank is $\dim {\cal K}$, then 
\begin{align*}
\Tr (\overline{\rho}_p - e^\lambda)P
\le \Tr (\overline{\rho}_p - e^\lambda)\{\overline{\rho}_p - e^\lambda \ge 0 \} .
\end{align*}
Thus, we obtain
\begin{align}
 \Tr \overline{\rho}_p P  
\le e^{\lambda} \dim {\cal K}+
\Tr \overline{\rho}_p \{\overline{\rho}_p - e^\lambda \ge 0 \}. \Label{h12}
\end{align} From (\ref{h11}), (\ref{L24}), (\ref{kore}) and (\ref{h12}),
\begin{align*}
& 1- \epsilon(F,D)= \sum_{i \in \Xi} p_i \Tr D(F(i))\rho_i \\
\le& \max\{ \Tr P \overline{\rho}_p' | P:\hbox{ projection on }
{\cal H}\otimes {\cal H}', \rank P= \dim {\cal K} \} \\
\le& \max\{ \Tr P \overline{\rho}_p | P:\hbox{ projection on }
{\cal H}, \rank P= \dim {\cal K} \} \\
\le &
e^{\lambda} \dim {\cal K}+
\Tr \overline{\rho}_p \{\overline{\rho}_p - e^\lambda \ge 0 \}.
\end{align*}
We obtain (\ref{L20}).
Since 
$\Tr \overline{\rho}_p \{\overline{\rho}_p - e^\lambda 
\,< 0 \} = 1- \Tr \overline{\rho}_p 
\{\overline{\rho}_p - e^\lambda
\ge 0 \}$,
the inequalities (\ref{L10}) and (\ref{L20}) hold.
Applying Markov inequality (\ref{Ma})
given in Appendix \ref{AMa}
to the probability ${\bf a}=\{a_i\}$ and 
the random variable $a_i^t$,
we obtain the inequality 
\begin{align*}
\Tr \overline{\rho}_p \{ \overline{\rho}_p - e^\lambda \ge 0 \}
\le e^{-t\lambda} \Tr {\overline{\rho}_p}^{1+t} \quad \forall t \ge 0,
\end{align*}
where $a_1,\ldots,a_d$ are eigenvalues of 
$\overline{\rho}_p$. 
Substituting 
$1+t$ for $s$,
we obtain (\ref{L21}).
\endproof
\begin{rem}
Assume that
$D$ is not a CP map but
a positive map.
In this case, 
the inequality
\begin{align}
1- \epsilon (F,D) \le 2 e^{\lambda} \dim {\cal K} +
2 \Tr \overline{\rho}_p \{ \overline{\rho}_p - e^\lambda \ge 0 \}
 \Label{L52} 
\end{align}
holds for $\forall \lambda \in \real$
instead of (\ref{L20}).
This inequality is proven in Appendix \ref{aped}.
\end{rem}
\section{Proof of the converse part of Theorem \ref{T20}}\Label{s6}
First, using Lemma \ref{L1}, 
we prove inequality
(\ref{12-3}).
\begin{align}
r_{e,V}(R|\{\rho_i,p_i\}_{i \in  \Xi})
\le  \max_{0 \le s \le 1} \frac{(1-s)R - \psi(s)}{s}\Label{s34}.\end{align}
Assume that a sequence of
visible codes
$\{ ({\cal K}_n,F_n,D_n)\}$ satisfies that
\begin{align}
\varlimsup \frac{1}{n}\log \dim {\cal K}_n  \le R  \Label{s316}.
\end{align}
It follows from (\ref{L10}) in Lemma \ref{L1} that
\begin{align*}
\epsilon(F_n,D_n)
\ge \Tr \overline{\rho}_p^{\otimes n} \{\overline{\rho}_p^{\otimes n} - e^{-nS}\le 0 \}
- e^{-nS} \dim {\cal K}_n.
\end{align*}
When $S-R \ge \eta(S):= \varlimsup  \frac{-1}{n}\log 
\Tr \overline{\rho}_p^{\otimes n} \{\overline{\rho}_p^{\otimes n} - e^{-nS}\le 0 \}$,
\begin{align*}
\varlimsup
-\frac{1}{n}
\log \epsilon(F_n,D_n)
\le \eta(S).
\end{align*}
Therefore, 
we have 
\begin{align*}
\varlimsup
-\frac{1}{n}
\log \epsilon(F_n,D_n)
&\le \inf\{\eta(S)| S-R \ge \eta(S)\} \\
&= \inf\{\eta(S)| S-\eta(S) \ge R\}.
\end{align*} 
Now,
applying (\ref{Cra4}) to the 
random variable 
$-\log a_i$ under the probability
distribution ${\bf a}$,
we obtain
\begin{align*}
\begin{array}{lll}
\eta(S)&= (1-s(S))-\psi(s(S)) & \hbox{ if }
H(\rho) \le S \le -\psi'(0) \\
\eta(S)&\ge \eta(-\psi'(0)) &\hbox{ if }
S \,>  -\psi'(0) \\
\eta(S)&=0  &\hbox{ if } S\,< H(\rho),
\end{array}
\end{align*}
where the definition of $s(S)$ 
is given in Lemma \ref{L11} in Appendix \ref{appen1}.
When $H(\rho) \,< S \,<-\psi'(0) $,
\begin{align*}
\frac{\,d \eta(S)}{\,d S}
&= 1- s(S) \ge 0 \\
\frac{\,d (S-R-\eta(S))}{\,d S}
&= s(S) \ge 0.
\end{align*}
When $H(\rho) \,< R \,< \log d =\psi(0)=
-\psi'(0)-\eta(-\psi'(0))$,
we obtain
\begin{align*}
\inf\{\eta(S)| S- \eta(S)\ge R\}
= \eta(S_R)= S_R-R.
\end{align*}
When $0 \le R \le H(\rho)$,
we obtain
\begin{align*}
\inf\{\eta(S)| S- \eta(S)\ge R\}
\le
\inf\{\eta(S)| S\ge H(\rho)\}=0.
\end{align*}
Using Lemma \ref{L9},
we obtain (\ref{12-3}).

Next, we prove (\ref{12-4}).
Assume that
a sequence of visible codes
$\{ ({\cal K}_n,F_n,D_n)\}$ satisfies that
\begin{align}
\varlimsup \frac{1}{n}\log \dim {\cal K}_n  \le R  \Label{s316.1}.
\end{align}
When $ H(\rho) \le R $,
it is trivial that
\begin{align*}
&\varliminf -\frac{1}{n}\log
(1- \epsilon(F_n,D_n)) \ge
S_{H(\rho)}-H(\rho)=0.
\end{align*} 
Lemma \ref{L9} implies (\ref{12-4}).

Assume that 
$a_1 = a_k \,> a_{k+1}$ and 
$\log k \,< R \,< H(\rho)$.
Since $\log \Tr (\overline{\rho}_p^{\otimes n})^{s}=
n \psi(s)$,
substituting $\lambda := -n S_R$ and 
$s:= s(S_R)\ge 1$ into (\ref{L21}),
we have
\begin{align}
1- \epsilon(F_n,D_n)
\le
e^{-n(S_R-R)} + e^{-n(S_R (1-s(S_R)) - \psi(s(S_R))) } \Label{s21}.
\end{align}
Note that the definitions of $S_R,s(S)$ are
given in Lemma \ref{L11}.
Since $S_R-R
= S_R (1-s(S_R)) - \psi(s(S_R)) $, 
we have
\begin{align}
&\varliminf -\frac{1}{n}\log
(1- \epsilon(F_n,D_n)) \ge 
S_R-R \nonumber \\ 
&=\frac{(1-s(S_{R})) R + \psi (s(S_{R}))}{s(S_{R})}\Label{s23},
\end{align} 
where the last inequality
follows from $S_R= \frac{R+\psi(s(S_R))}{s(S_R)}$
obtained from (\ref{s323}). From Lemma \ref{L9}, 
we obtain (\ref{12-4}).

Assume that $0\le R \le \log k$.
Substituting $\lambda:= -n ( - \log a_1 - \epsilon)$
into (\ref{L21}),
we have
\begin{align}
&1- \epsilon(F_n,D_n)\nonumber \\
\le &
e^{-n(- \log a_1 - \epsilon -R)} + 
e^{-n((-\log a_1 -\epsilon) (1-s) 
- \psi(s) ) } \Label{hh21}
\end{align}
for $\forall \epsilon \,> 0$ and $\forall 
s \ge 1$.
Since
\begin{align*}
&\lim_{s \to \infty}
(-\log a_1 -\epsilon) (1-s) 
- \psi(s) \\
= &\lim_{s \to \infty}
\epsilon(s-1) - 
\log \frac{k a_1^s}{\sum_{i=1}^d a_i^s}
- \log a_1 + \log k = \infty,
\end{align*}
we have
\begin{align*}
\varliminf -\frac{1}{n}\log
(1- \epsilon(F_n,D_n)) \ge 
- \log a_1 -\epsilon -R.
\end{align*}
Arbitrarity of $\epsilon \,> 0$
implies 
\begin{align*}
\varliminf -\frac{1}{n}\log
(1- \epsilon(F_n,D_n)) \ge 
- \log a_1 -R.
\end{align*}
Lemma \ref{L9} implies (\ref{12-4}).

\section{Discussion}
When the source $\rho_i$ is mixed and 
has no trivial redundancies,
Koashi and Imoto \cite{KI}
proved that the bound $R_{B}$
equals $H(\rho)$ in the blind case.
Lemma \ref{L3} holds for the mixed case.
However, its optimality is not proven in the sense
of exponents in the mixed case.
In this case it may not be optimal.

It is interesting that our exponent corresponds to
the exponents of 
the variable-length universal entanglement 
concentration given by Hayashi and 
Matsumoto\cite{HM}
and the fixed-length entanglement 
concentration given by Hayashi et. al.\cite{HKMMW}.
However, our error exponent corresponds to the 
success exponent of \cite{HM},
and our fidelity exponent
corresponds to the failure exponents of \cite{HM}
and \cite{HKMMW}.
Note that in \cite{HKMMW} the optimal exponent $r$
is given as the function of the rate $R$
while in this paper and \cite{HM}, 
the rate $R$ is given as a function the optimal exponent $r$.
In addition, in quantum hypothesis testing,
an error exponent similar to 
(\ref{T2}) is given in Ogawa and 
Hayashi\cite{O-H}.
\section*{Acknowledgment}
The author is grateful to Dr. A. Winter for
advice on Nielsen and Kempe's paper \cite{NK}.
He acknowledges stimulating discussions with 
Professor H. Nagaoka, Professor K. Matsumoto and
Dr. T. Ogawa.
He, also, thanks an anonymous referee for
useful comments.
\appendix 
\section{Equivalence between different characterizations}\Label{appen1}
In the classical case, the exponent has two forms 
\cite{CK}\cite{Blahut}\cite{CL}.
Following Ogawa and Nagaoka \cite{Oga-Nag:test},
we prove this equivalence in the quantum source coding case.
In this section 
we treat a state 
$\rho:= \sum_i a_i | e_i \rangle \langle e_i |$,
and the function
$\psi(s):= \log \Tr \rho^s$,
where $a_i \ge a_{i+1}$.
We assume that $a_1 = a_k \,> a_{k+1}$
and $d = \dim {\cal H}$.
\begin{lem}\Label{L11}
If $-\log a_1 \,< S \le - \psi'(0)$
and  $\log k \,< R \,< \log d$,
we can uniquely define $s(S) \ge 0$ 
and $S_R$
such that
\begin{align}
S&= - \psi'(s(S)), \Label{s324}\\
R&= s(S_R)S_R +\psi(s(S_R)).\Label{s323}
\end{align}
Conversely, when $R \le \log k$,
\begin{align}
R\,< - s \psi'(s) +\psi(s).\Label{hh11}
\end{align}
\end{lem}
\begin{proof}
Since
\begin{align}
\psi''(s)= \frac{\Tr (\log \rho)^2 \rho^s
\Tr \rho^s - \left(\Tr (\log \rho) \rho^s\right)^2}
{\left(\Tr \rho^s\right)^2} \,> 0 \Label{b2}
\end{align}
for $s \,>0$, the function
$- \psi'(s)$ is monotone decreasing.
Because $\lim _{s \to \infty} 
- \psi'(s)= \log a_1$,
$s(S)$ is uniquely defined in 
$(-\log a_1, - \psi'(0)]$.

When $S \in (-\log a_1 , - \psi'(0)]$,
we can calculate
\begin{align*}
\frac{\,d}{\,d S}
s(S)S +\psi(s(S))=s(S) \,> 0.
\end{align*}
As shown latter, the equation
\begin{align}
\lim _{s \to \infty}
-\psi'(s )s +\psi(s) = \log k . \Label{12-09}
\end{align}
holds. Since
\begin{align*}
-\psi'(0)0
 +\psi(0)= \psi(0)= d ,
\end{align*}
$S_R$ also is uniquely defined.
The inequality
$\frac{\,d }{\,d s}
(- s \psi'(s) +\psi(s))
= -s \psi''(s) \le 0$ yields (\ref{hh11}).

Finally, we show (\ref{12-09}).
We calculate as
\begin{align*}
&-\psi'(s )s +\psi(s) 
=\sum_{i=1}^d -s \log a_i \frac{a_i^s}
{\sum_{j=1}^d a_j^s}
+ \log \sum_{i=1}^d a_i^s \\
=&
- \sum_{i=k+1}^d s \frac{a_i^s}{\sum_{j=1}^d a_j^s}
\log a_i 
+ \log \sum_{j=1}^d a_j^s
- \log k a_1^s \\
&\quad + \left( -k s \frac{a_1^s}{\sum_{j=1}^d a_j^s}
\log a_1 
+ s \log a_1 \right) 
+ \log k
\\
=&
- \sum_{i=k+1}^d s 
\frac{a_i^s}{\sum_{j=1}^d a_j^s}\log a_i 
+ \log \frac{\sum_{j=1}^d a_j^s}{k a_1^s} \\
&\quad + 
s \frac{\sum_{i=k+1}^d a_i^s}{\sum_{j=1}^d a_j^s}
\log a_1  
+ \log k.
\end{align*}
The terms $\frac{a_i^s}{\sum_{j=1}^d a_j^s}$
and $\frac{\sum_{i=k+1}^d a_i^s}
{\sum_{j=1}^d a_j^s}$
exponentially go to $0$ as $s \to \infty$.
The term $\frac{\sum_{j=1}^d a_j^s}{k a_1^s}$
goes to $1$.
Thus, we obtain (\ref{12-09}).

\end{proof}

\begin{lem}\Label{L9}
When $\log k \,< R \,< \log d$,
the equations
\begin{align}
 S_R-R 
&=  S_R (1-s(S_R)) - \psi(s(S_R)) \Label{kk-1}\\
&=
\frac{(1-s(S_{R})) R - 
\psi (s(S_{R}))}{s(S_{R})} 
\Label{kk} \\
&= \min_{H({\bf b}) = R}
D({\bf b}\|{\bf a}) =
\min_{H(\sigma) = R}
D(\sigma\|\rho)  \Label{kk2}
\end{align}
hold,
where $\sigma$ is a state on ${\cal H}$ and
${\bf b}$ is a probability on $\{ 1, \ldots, d\}$.
When $0 \le R \le \log k$, the equations
\begin{align}
\min_{H({\bf b}) = R}
D({\bf b}\|{\bf a}) =
\min_{H(\sigma) = R}
D(\sigma\|\rho)  
=-\log a_1 - R . \Label{hh1}
\end{align}
hold.
When $H(\rho) \,< R \,< \log d$,
\begin{align}
 S_R-R 
&= \min_{H({\bf b}) \ge R}
D({\bf b}\|{\bf a}) =
\min_{H(\sigma) \ge R}
D(\sigma\|\rho)  \Label{hh2}\\
&=
\max_{0 \,< s \le 1}
\frac{(1-s)R - \psi(s)}{s} \Label{kk4}\\
0&=\min_{H({\bf b}) \le R}
D({\bf b}\|{\bf a}) =
\min_{H(\sigma) \le R}
D(\sigma\|\rho)  \Label{hh5}\\
&= \max_{s \ge 1}
\frac{(1-s)R - \psi(s)}{s}.\Label{hh6}
\end{align}
When $\log k \,< R \,< H(\rho)$,
\begin{align}
0&= \min_{H({\bf b}) \ge R}
D({\bf b}\|{\bf a}) =
\min_{H(\sigma) \ge R}
D(\sigma\|\rho)   \Label{hh3} \\
& =
\max_{0 \,< s \le 1}
\frac{(1-s)R - \psi(s)}{s} \Label{hh4}\\
 S_R-R &= \min_{H({\bf b}) \le R}
D({\bf b}\|{\bf a}) =
\min_{H(\sigma) \le R}
D(\sigma\|\rho)  \Label{hh7} \\
&=
\max_{s \ge 1}
\frac{(1-s)R - \psi(s)}{s}. \Label{kk5}
\end{align}
When $0 \le R \le \log k$,
\begin{align}
0&= \min_{H({\bf b}) \ge R}
D({\bf b}\|{\bf a}) =
\min_{H(\sigma) \ge R}
D(\sigma\|\rho)   \Label{hh32} \\
& =
\max_{0 \,< s \le 1}
\frac{(1-s)R - \psi(s)}{s} \Label{hh42}\\
\log a_1 -R 
&= \min_{H({\bf b}) \le R}D({\bf b}\|{\bf a}) =
\min_{H(\sigma) \le R}
D(\sigma\|\rho)   \Label{hh8} \\
&= \sup_{s \ge 1}
\frac{(1-s)R - \psi(s)}{s}. \Label{hh9}
\end{align}
\end{lem}
\begin{proof} 
Equation (\ref{kk-1}) follows from (\ref{s323}).
Equation (\ref{s323}) yields
\begin{align*}
S_R= \frac{R- \psi(s(S_R))}{s(S_R)}.
\end{align*}
Substituting the above equation into 
$S_R-R$,
we obtain (\ref{kk}).
We prove (\ref{kk2}).
Assume that $\log k \,< R \,< \log d$.
Letting $\rho_s:= \frac{\rho^s}{\Tr \rho^s}$,
we calculate
\begin{align*}
& D(\sigma \| \rho )-D(\rho_s\|\rho) \\
=& \Tr  \sigma (\log \sigma - \log \rho) 
-\Tr  \frac{\rho^s}{\Tr \rho^s}
\left(\log \left( \frac{\rho^s}{\Tr \rho^s }\right)
- \log \rho\right)\\
=& \Tr \sigma \left( \log \sigma 
- \log \left(\frac{\rho^s}{\Tr \rho^s }\right)\right) \\
&\quad + \Tr
\left( \sigma - \left( \frac{\rho^s}{\Tr \rho^s }\right)\right)
\left( \log \left( \frac{\rho^s}{\Tr \rho^s }\right)
- \log \rho \right) \\
=& D(\sigma \|\rho_s) -
(1-s) \Tr
\left( \sigma - \left( \frac{\rho^s}{\Tr \rho^s }\right)\right)
\log \rho  \\
&-H(\sigma)+H(\rho_s)  \\
=&
\Tr \sigma \left( \log \sigma - \log \left( \frac{\rho^s}{\Tr \rho^s}
\right)\right) \\
&\quad +\Tr 
\left( \sigma - \left( \frac{\rho^s}{\Tr \rho^s }\right)\right)
\log \left( \frac{\rho^s}{\Tr \rho^s }\right)\\
=&  D(\sigma\|\rho_s) + s \Tr
\left( \sigma - \left( \frac{\rho^s}{\Tr \rho^s }\right)\right)
\log \rho.
\end{align*}
Equation (\ref{s323}) guarantees that
$
H(\rho_{s(S_R)})=R .
$
Assuming that $H(\sigma)=R$,
we have
\begin{align*}
&\frac{D(\sigma\|\rho_{s(S_R)})}{s(S_R)}
=
- \Tr \left( \sigma - \rho_{s(S_R)}
\right)
\log \rho \\
=&
\frac{1}{1-s(S_R)}
\left( D(\sigma\|\rho)-D(\rho_{s(S_R)}\|\rho)
- D(\sigma\|\rho_{s(S_R)})
\right)
\end{align*}
i.e.,
\begin{align*}
 D(\sigma\|\rho)-D(\rho_{s(S_R)}\|\rho)
=
\frac{1}{s(S_R)}
D(\sigma\|\rho_{s(S_R)}) \ge 0.
\end{align*}
It implies that
\begin{align*}
D(\rho_{s(S_R)}\|\rho)
= \min_{H(\sigma)=R} D(\sigma\|\rho)
=\min_{H({\bf b})=R} D({\bf b}\|{\bf a}).
\end{align*}
Note that $\rho_s$ is commutative with $\rho$.
Equation (\ref{s323}) yields 
\begin{align*}
D(\rho_{s(S_R)}\|\rho)
=& \psi'(s(S_R))( 1 - s(S_R)) - \psi(s(S_R)) \\
=& S_R( 1 - s(S_R)) - \psi(s(S_R)).
\end{align*}
Then,
we obtain (\ref{kk2}). 

Next, we proceed (\ref{hh1}) and
assume that $0\le R \le \log k$.
When $H(\sigma)=R$,
\begin{align*}
& D(\sigma\|\rho)=
\Tr \sigma \log \sigma + \Tr
\sigma (-\log \rho) \\
\ge& - H(\sigma) + \Tr \sigma (- \log a_1)
= - \log a_1 - R.
\end{align*}
Let ${\bf c}:=\{ c_i \}_{i=1}^k$ be
a probability whose entropy is $R$.
Then we have
\begin{align*}
D\left(\left. \sum_{i=1}^k c_i |e_i \rangle 
\langle 
e_i|\right\|\rho\right)
&= \sum_{i=1}^k c_i (\log c_i - \log a_1) \\
&= - \log a_1 -R.
\end{align*}
Thus, we obtain (\ref{hh1}),
which implies (\ref{hh8}).

Taking the derivative with respect to $R$ in
(\ref{s323}), we have
\begin{align}
\frac{\,d }{\,d R}s(S_R)
=\frac{-1}{s(S_R)\psi''(s(S_R))} \,< 0. \Label{h9}
\end{align} From (\ref{s324}),
we have
\begin{align*}
\frac{\,d }{\,d R}
(S_R- R) & =
- \psi''(s(S_R))\frac{\,d }{\,d R}s(S_R)
-1
= \frac{1- s(S_R)}{s(S_R)}. \\
\frac{\,d }{\,d R}
(S_R- R) & =
\frac{1}{s^3(S_R)}{\psi''(s(S_R))} \,> 0.
\end{align*}
Thus, the function $R \mapsto S_R-R$
is convex, and 
$\frac{\,d }{\,d R}
(S_R- R)=0$ if and only if
$s(S_R)=1$, i.e. $R=H(\rho)$.
The function takes minimum value $0$ 
at $R=H(\rho)$
because $S_{H(\rho)}-H(\rho)=0$.
Therefore, we obtain 
(\ref{hh2}),
(\ref{hh5}),
(\ref{hh3}), 
(\ref{hh7}),
and 
(\ref{hh32}).

Next, we discuss the other
forms described by $\psi$.
We can calculate the derivatives as
\begin{align}
\frac{\,d }{\,d s}
\frac{(1-s)R - \psi(s)}{s}
&=
\frac{-R - s \psi'(s) +\psi(s)}{s^2} 
\Label{hh10} \\
\frac{\,d }{\,d s}(-R - s \psi'(s) +\psi(s))
&=-s \psi''(s) \le 0 \Label{b1},
\end{align}
where the last inequality follows from 
(\ref{b2}).
In (\ref{b1}) and (\ref{b2}),
the equalities hold
if and only if $s=0$.

Assume $\log k \,< R \,< \log d$.
Since it follows from
(\ref{s324}) and (\ref{s323})
that
\begin{align}
-R - s(S_R) \psi'(s(S_R)) +\psi(s(S_R))=0, \Label{h8}
\end{align}
the equation
\begin{align*}
\max_{s\,> 0 }
\frac{(1-s)R - \psi(s)}{s}
=
\frac{(1-s(S_{R})) R + \psi (s(S_{R}))}{s(S_{R})} 
\end{align*}
holds.
Relation (\ref{h9}) implies that 
the function $R \mapsto s(S_R)$
strictly monotonically decreases,
and $s(S_R) \ge 1$ if and only if $R \le H(\rho)$.
Therefore,
\begin{align*}
&\max_{0 \,< s \le 1}\frac{(1-s)R - \psi(s)}{s}
\\
=&
\left\{
\begin{array}{ll}
\frac{(1-s(S_{R})) R + \psi (s(S_{R}))}{s(S_{R})} 
& \hbox{ if } H(\rho) \,< R \,< \log d\\
0 
&\hbox{ if } \log k \,< R \le H(\rho) 
\end{array}\right.
\\
&\max_{s \ge 1}\frac{(1-s)R - \psi(s)}{s}\\
=&
\left\{
\begin{array}{ll}
0 & \hbox{ if } H(\rho) \,< R \,< \log d\\
\frac{(1-s(S_{R})) R + \psi (s(S_{R}))}{s(S_{R})}
&\hbox{ if } \log k \,< R \le H(\rho) 
\end{array}\right.
\end{align*}
Note that
$\frac{(1-1)R - \psi(1)}{1}=0$.
We obtain (\ref{kk4}), (\ref{hh6}), (\ref{hh4}) and
(\ref{kk5}).

When $0 \le R \le \log k$,
Lemma \ref{L11} guarantees that
the RHS of (\ref{hh10}) is positive for any $s
\,> 0$.
Thus,
\begin{align*}
\sup_{s\,> 0 }
\frac{(1-s)R - \psi(s)}{s}
&= \lim_{s \to \infty}\frac{(1-s)R - \psi(s)}{s}
\\
&= - \log a_1 -R,
\end{align*}
which implies 
\begin{align*}
\max_{0 \,< s \le 1}\frac{(1-s)R - \psi(s)}{s}
& = 0 \\
\sup_{s \ge 1}\frac{(1-s)R - \psi(s)}{s}
&= - \log a_1 -R.
\end{align*}
We obtain (\ref{hh4}) and (\ref{hh9}).
\end{proof}
\section{Representation theoretical type methods}\Label{appen2}
In this section, 
we prove the following two lemmas
used in our proof of Lemma \ref{L3}.
We assume that
$\rho= \sum_{i=1}^d a_i |e_i \rangle
\langle e_i|$ and $d$ is the dimension of ${\cal H}$.
\begin{lem}\Label{L110}
The relations
\begin{align}
& \exp \left( n H\left(\frac{\bf n}{n}\right)\right) 
(n+d)^{- \frac{d(d+1)}{2}}\nonumber\\
\le
& \dim {\cal V}_{{\bf n}} \Label{a4-1}\\
\le &(n+d)^{2d}
\exp \left( n H\left(\frac{\bf n}{n}\right)\right) \Label{a4} \\
\# \{ {\bf n} | {\bf n} \in Y_n\}
\le& (n+1)^d\Label{h20} \\
\dim {\cal U}_{\bf n} \le &(n+1)^d \Label{h21}
\end{align}
hold, where $C({\bf n})$ is defined as
\begin{align*}
C({\bf n}):= 
\frac{n!}{n_1! n_2! \ldots n_d !}.
\end{align*}
\end{lem}
\begin{proof}
Inequality (\ref{h20})
is trivial. 
Using Young index ${\bf n}$,
the basis of ${\cal U}_{\bf n}$
is described by
$\{ e_{{\bf n}'} \}_{{\bf n}' \in Y^{\bf n}}$,
where the set $Y^{\bf n}$ is defined as
\begin{align*}
Y^{\bf n} 
:= \left\{
{\bf n}'= \{ n'_i\} \in \mathbb{Z}^d\left|
\begin{array}{l}
\sum_i n'_i=\sum_i n_i, \\
\sum_{i=1}^m n'_{s(i)} \le \sum_{i=1}^m
n_i, \\
1 \le \forall m \le d-1, \\
s \hbox{ is any permutation} 
\end{array}
\right.\right\}.
\end{align*}
Thus, we obtain (\ref{h21}).
Note that
the correspondence
${\bf n}'$ and $e_{{\bf n}'}$
depends on the choice of
Cartan subalgebra, i.e.
the choice of basis of ${\cal H}$.

According to Weyl \cite{Weyl}, and Iwahori \cite{Iwa},
the following equation holds
and is evaluated as:
\begin{align}
&\dim {\cal V}_{{\bf n}} \nonumber \\
=& \frac{n !}
{(n_1+d-1)! (n_2+d-2)! \ldots n_d!}
\prod_{j \,> i}(n_i-n_j-i+j) \nonumber \\
\le& \frac{n !}{n_1! n_2! \ldots n_d!}
\prod_{j \,> i}(n_i-n_j-i+j)\nonumber \\
\le& C({\bf n})(n+d)^{2d} \Label{a4-2}\\
\le& (n+d)^{2d}
\exp \left( n H\left(\frac{\bf n}{n}\right)\right). \nonumber 
\end{align}
Thus, we obtain (\ref{a4}).
As an opposite inequality, we have
\begin{align*}
&\dim {\cal V}_{{\bf n}}\\
 \ge &
\frac{n !}{(n_1+d-1)! (n_2+d-2)! \ldots n_d!} \\
 \ge &
\frac{n !}{n_1! n_2 ! \ldots n_d!}
\left(\frac{1}{n+d}\right)^{d-1}
\left(\frac{1}{n+d}\right)^{d-2}
\cdots \left(\frac{1}{n+d}\right)^{0} \\
=&
C({\bf n})
\left(\frac{1}{n+d}\right)^{\frac{d(d-1)}{2}}
\ge 
\exp \left( n H\left(\frac{\bf n}{n}\right)\right) 
(n+d)^{- \frac{d(d+1)}{2}},
\end{align*}
where the last inequality follows from
\begin{align*}
C({\bf n}) \ge \frac{1}{(n+1)^d}
\exp \left( n H\left(\frac{\bf n}{n}\right)\right) ,
\end{align*}
which is easily proven by the type 
method \cite{CK}.
We obtain (\ref{a4-1}).
\end{proof}
The following is essentially equivalent to
Keyl and Werner's result \cite{KW}.
For the reader's convenience, we give a simpler proof.
\begin{lem}\Label{lee}
The following relations
\begin{align}
&(n+d)^{-\frac{d(d+1)}{2}}
\exp \left( -n 
D\left(\left. \frac{\bf n}{n} 
\right\|{\bf a}\right)\right)\nonumber \\
\le &
\Tr P_{{\bf n}} \rho^{\otimes n} \Label{f-e31}\\
\le &
(n+d)^{3d }
\exp \left( -n 
D\left(\left. \frac{\bf n}{n} \right\|{\bf a}\right)\right)
\Label{e31} \\
&(n+d)^{-\frac{d(d+1)}{2}}
\exp \left( -n 
\min_{{\bf n} \in n {\cal R}\cap Y_n}
D\left(\left. \frac{\bf n}{n} \right\|{\bf a}\right)
\right) \nonumber \\
\le &
\sum_{
\frac{\bf n}{n} \in {\cal R}}
\Tr P_{{\bf n}} \rho^{\otimes n} \Label{f-e32}\\
\le &
(n+d)^{4d }
\exp \left( -n 
\inf_{{\bf b} \in {\cal R}}
D( {\bf b} \|{\bf a})
\right), 
\Label{e32}
\end{align}
hold,
where ${\cal R}$ is 
a subset consisting of probabilities
on $\{ 1, \ldots, d \}$
and we denote
the projection to ${\cal W}_{\bf n}$ by
$P_{\bf n}$.
\end{lem}
\begin{proof}
Let ${\cal U}_{\bf n}'$ be an irreducible
representation of $SU(d)$ in ${\cal H}^{\otimes n}$,
which is equivalent to ${\cal U}_{\bf n}$.
We denote its projection by $P_{\bf n}'$.
Now, we choose the basis 
$\{ e_{{\bf n}'}\}_{{\bf n}' \in Y^{\bf n}}$
of ${\cal U}_{\bf n}'$
depending the basis $\{e_i\}$ of ${\cal H}$.
The base $e_{{\bf n}'}$ is the eigenvector 
of $\rho^{\otimes n}$ with the
eigenvalue $\prod_{i=1}^d a_i^{n_i'}$.
Since ${\bf n}'$ is majorized by ${\bf n}$,
we can calculate the operator norm by
\begin{align}
\left \|
P_{\bf n}' \rho^{\otimes n}
P_{\bf n}' 
\right\| = \prod_{i=1}^d a_i^{n_i},\Label{a5}
\end{align}
where $\| X \| := \sup_{x \in {\cal H}}\| X x \|$.
from (\ref{h21}), (\ref{a4-2}) and (\ref{a5}), the relations
\begin{align*}
\Tr P_{{\bf n}} \rho^{\otimes n}
&=
\dim {\cal V}_{\bf n} \times 
\Tr
P_{\bf n}' \rho^{\otimes n}
\le  
(n+d)^{3d} C({\bf n}) \prod_{i=1}^d a_i^{n_i}\\
&= (n+d)^{3d} {\rm Mul}( {\bf a}, {\bf n} )  
\end{align*}
hold, where 
we denote the multinomial distribution of ${\bf a}$ by
${\rm Mul}( {\bf a}, \bullet )$.
Inequality (\ref{h20}) guarantees 
\begin{align*}
\frac{1}{(n+1)^d}\exp \left(-n D\left(\left.\frac{\bf n}{n}\right\|
{\bf a}\right)\right) &\le
{\rm Mul}( {\bf a}, {\bf n}) \\
&\le
\exp \left(-n D\left(\left.\frac{\bf n}{n}\right\|
{\bf a}\right)\right) .
\end{align*}
Thus, we obtain inequality (\ref{e31}).
Inequality (\ref{h20}) guarantees that
\begin{align*}
\sum_{
{\bf n} \in n {\cal R}\cap Y_n }
\Tr P_{{\bf n}} \rho^{\otimes n}
\le
(n+d)^{4d}
\exp \left( -n 
\inf_{{\bf b} \in {\cal R}}
D( {\bf b} \|{\bf a})
\right),
\end{align*} 
which implies inequality (\ref{e32}). From 
(\ref{a5}), we have
\begin{align*}
&\Tr P_{\bf n}\rho^{\otimes n}
= \dim {\cal V}_{\bf n} \Tr P_{\bf n}'\rho^{\otimes n} \\
\ge &
\exp \left( n H\left(\frac{\bf n}{n}\right)\right) 
(n+d)^{- \frac{d(d+1)}{2}}
\prod_{i=1}^d a_i^{n_i} \\
= &
(n+d)^{- \frac{d(d+1)}{2}}
\exp \left( -n 
D\left(\left. \frac{\bf n}{n}\right\|
{\bf a} \right)\right) .
\end{align*}
Therefore, we obtain inequalities 
(\ref{f-e31}) and (\ref{f-e32}).
\end{proof}
\section{Proof of (\ref{h17}), (\ref{h16}), (\ref{h27}) and (\ref{h26})}
\Label{apec}
Since 
\begin{align*}
\epsilon(F,D)&= \sum_{i \in \Xi} p_i 
(1 - \Tr D \circ F(i) \rho_i) \\
&\ge
\sum_{i \in \Xi} p_i 
(1 - \sqrt{\Tr D \circ F(i) \rho_i})
=\epsilon_b(F,D),
\end{align*}
the inequalities
\begin{align}
r_{e,V,b}(R|\{\rho_i,p_i\}_{i \in  \Xi})
& \ge r_{e,V}(R|\{\rho_i,p_i\}_{i \in  \Xi}) \Label{h15} \\
r_{e,V}^*(R|\{\rho_i,p_i\}_{i \in  \Xi})
& \ge r_{e,V,b}^*(R|\{\rho_i,p_i\}_{i \in  \Xi}) \nonumber
\end{align}
hold.
Similarly, we can prove that
\begin{align*}
r_{e,B,b}(R|\{\rho_i,p_i\}_{i \in  \Xi})
& \ge r_{e,B}(R|\{\rho_i,p_i\}_{i \in  \Xi}) \\
r_{e,B}^*(R|\{\rho_i,p_i\}_{i \in  \Xi})
& \ge r_{e,B,b}^*(R|\{\rho_i,p_i\}_{i \in  \Xi}).
\end{align*}
Using Jensen's inequality, we have
\begin{align*}
\epsilon(F_n,D_n)
&=
\sum_{i \in \Xi} p_i 
(1 - \Tr D_n \circ F_n(i) \rho_i) \\
&\le 1- 
\left(\sum_{i \in \Xi} p_i 
\sqrt{\Tr D_n \circ F_n(i) \rho_i}\right)^2 \\
& 
= 1- \left(1- \epsilon_b(F_n,D_n)\right)^2 
\le 2 \epsilon_b(F_n,D_n)
\end{align*}
Thus, we obtain the opposite inequality 
from (\ref{h15})
and then obtain (\ref{h16}).
Similarly, we can prove (\ref{h17}).
\section{Proof of (\ref{L52})}\Label{aped}
For any visible code $({\cal K},F,D)$,
we define 
an operator $T$
by $T
:= \{ D(\Id) -1 \le 0\} D(\Id) \{ D(\Id) -1 \le 0\}+
\{ D(\Id) -1 \,> 0\}$.
The operator inequality
\begin{align}
P\rho P+ (\Id-P) \rho (\Id-P) \ge \frac{1}{2} \rho \Label{opi}
\end{align}
holds for any projection $P$.
It is sufficient for (\ref{opi}) to show the pure state case.
The pure state case of (\ref{opi}) is
directly proven using the inequality 
$2(|x|^2 + |y|^2) \ge |x+y|^2$ for any two complex numbers $x,y$.
Therefore, 
\begin{align}
&\{ D(\Id) -1 \le 0\}D(F(i))\{ D(\Id) -1 \le 0\}
\nonumber\\
&\quad + \{ D(\Id) -1 \,> 0\}D(F(i))\{ D(\Id) -1 \,> 0\}\nonumber\\
\ge& \frac{1}{2} D(F(i))\Label{2}.
\end{align}
The inequality 
$ D(\Id)  \ge D(F(i))$ 
follows from the inequality $\Id \ge F(i)$.
Thus, 
\begin{align}
&\{ D(\Id) -1 \le 0\}D(I)\{ D(\Id) -1 \le 0\}
\nonumber\\
\ge &
\{ D(\Id) -1 \le 0\}D(F(i))\{ D(\Id) -1 \le 0\}\Label{3}.
\end{align}
\par From the relations 
$\Tr D(F(i))=1$ and $D(F(i)) \ge 0$,
we can prove 
\begin{align}
&\{ D(\Id) -1 \,> 0\}\nonumber\\ 
\ge &
\{ D(\Id) -1 \,> 0\}D(F(i))\{ D(\Id) -1 \,> 0\} \Label{4}.
\end{align}
It follows from (\ref{3}) and (\ref{4}) that
\begin{align}
&\{ D(\Id) -1 \le 0\} D(\Id) \{ D(\Id) -1 \le 0\}+
\{ D(\Id) -1 \,> 0\} \nonumber \\
\ge &
\{ D(\Id) -1 \le 0\}D(F(i))\{ D(\Id) -1 \le 0\}
\nonumber \\
&\quad + \{ D(\Id) -1 \,> 0\}D(F(i))\{ D(\Id) -1 \,> 0\}\Label{5.1}.
\end{align} From (\ref{5.1}) and (\ref{2}), we have
\begin{align}
T \ge \frac{1}{2} D(F(i)). \Label{5}
\end{align}
Note that 
\begin{align}
\Tr T \le \Tr D(\Id) = \dim {\cal K} \Label{6}.
\end{align}
Since $\Id \ge T \ge 0$, we have
\begin{align}
& \Tr (\rho- e^\lambda )T \le
\Tr (\rho- e^\lambda ) \{ \rho- e^\lambda \ge 0 \}\nonumber \\
\le & \Tr \rho \{ \rho- e^\lambda \ge 0 \} 
\Label{7}.
\end{align} From (\ref{5}), (\ref{6}) and (\ref{7}),
we obtain (\ref{L20}).

\section{Markov inequality and
Cram\'{e}r's Theorem}\Label{AMa}
In this section,
we summarize 
Markov inequality and Cram\'{e}r's Theorem
which are applied in this paper.
Let $p$ be a probability distribution 
and $X$ be a positive real valued random variable.
For any real number $c \,>0$, we can easily 
prove the inequality
\begin{align}
\frac{E_p(X)}{c} \ge  p \{ X \ge c\}, \Label{Ma}
\end{align}
where $E_p$ presents the expectation 
under the distribution $p$.
This inequality is called {\it Markov inequality}.

This inequality can be used for large deviation 
evaluation as follows.
Let $Y$ be a real valued random variable.
In the $n$-i.i.d. setting, we focus on 
the random variable.
\begin{align*}
Y^n:= \sum_{i=1}^n \frac{Y_i}{n},
\end{align*}
where $Y_i$ is the $i$-th random variable
identical to $Y$.
Applying Markov inequality
for the random variable
$e^{tY^n}$, we have
\begin{align*}
p^n\{ Y^n \ge x \}
= p^n\{ e^{ntY^n} \ge e^{ntx} \}
\le
\frac{e^{n \phi(t)}}{e^{tx}}
\end{align*}
for $t \ge 0$,
which is equivalent to
\begin{align*}
\frac{-1}{n}
\log p^n\{ Y^n \ge x \}
\ge tx - \phi(t),
\end{align*}
where $\phi(t):= \log E_P(\exp (tY))$.
Therefore,
\begin{align}
\frac{-1}{n}
\log p^n\{ Y^n \ge x \}
\ge \sup_{t \ge 0} (tx - \phi(t) )\Label{Cra2}.
\end{align}
Conversely, 
the inequality 
\begin{align}
\lim \frac{-1}{n}
\log p^n\{ Y^n \,> x \}
\le \inf_{x' \,> x}I(x)
\Label{Cra1}
\end{align}
holds,
where
$I(x):= \sup_{t \in \real} (t x - \phi(t))$.
For a proof of (\ref{Cra1}), 
see Chapter II of Bucklew\cite{Buck}.
The pair of (\ref{Cra2}) and (\ref{Cra1})
is called {\it Cram\'{e}r's Theorem}.

In the following,
we discuss the case $\phi(t)$ is convex and
differentiable.
We define three real numbers $x_1,x_2$ and $x_3$ as
\begin{align*}
x_1:= \lim_{t \to \infty}\phi'(t) ,\quad
x_2:= \lim_{t \to -\infty}\phi'(t) ,\quad
x_3:= \phi'(0).
\end{align*}
For any $x \in (x_2,x_1)$, we can uniquely 
define $t(x)$ as
\begin{align*}
x= \phi'(t(x)).
\end{align*}
Then, 
\begin{align*}
I(x)&= x t(x)- \phi(t(x)) , \quad
I'(x)= t(x) , \\
I''(x)&= t'(x)= \frac{1}{\phi''(t(x))},
\end{align*}
where the last equation follows from 
\begin{align*}
1= \frac{\,d x}{\,d x}=
t'(x)\phi''(t(x)).
\end{align*}
Thus, we obtain 
\begin{align}
& \lim_{n \to \infty}
\frac{-1}{n} \log p^n
\{ Y_n \ge x \} \nonumber \\
= &
\left\{
\begin{array}{ll}
x t(x)- \phi(t(x)) & \hbox{ if }
x_3 \le x \le x_1 \\
+ \infty & \hbox{ if } x \,> x_1 \\
\phi(0)=0 &  \hbox{ if } x \,< x_3
\end{array}
\right.
\Label{Cra4}
\end{align}
except for $x=x_1$.

\end{document}